\documentclass[floats,floatfix,showpacs,amssymb,prd,twocolumn,superscriptaddress,nofootinbib,nolongbibliography,reprint]{revtex4-2}

\usepackage{amssymb,amsmath,verbatim,mathtools,needspace,enumitem,etoolbox,graphicx,physics,microtype,afterpage,bigints,gensymb,tabularx}

\usepackage[dvipsnames, usenames]{xcolor}
\definecolor{linkcolor}{rgb}{0.0,0.3,0.5}
\definecolor{dodgerblue}{HTML}{1E90FF}
\usepackage[unicode, colorlinks=true, linkcolor=linkcolor, citecolor=linkcolor, filecolor=linkcolor,urlcolor=linkcolor, pdfusetitle]{hyperref}
\usepackage[all]{hypcap}
\usepackage[T1]{fontenc}
\usepackage[utf8]{inputenc}
\usepackage{orcidlink}
\usepackage{amsmath,amsfonts,amsthm,bm}
\usepackage{booktabs}
\usepackage{multirow}
\usepackage{siunitx,soul}

\usepackage{makecell}

\newcommand{\ssim}{\mathchar"5218\relax\,}

\makeatletter
\newcommand*{\balancecolsandclearpage}{\close@column@grid \cleardoublepage \twocolumngrid}
\makeatother

\usepackage{lipsum}

\def\apj{{Astrophys.~J.}}\def\apjl{{Astrophys.~J.~Lett.}}
\def\apjs{{Astrophys.~J.~Supp.}}

\def\mnras{{Mon.~Not.~R.~Astron.~Soc.}}

\def\prd{{Phys.~Rev.~D}}

\def\prx{{Phys.~Rev.~X}}
\def\prl{{Phys.~Rev.~Lett.}}
\def\prr{{Phys.~Rev.~Res.}}

\def\lrr{{Living Rev. Relativ.}}
\def\cqg{{Classical~Quant.~Grav.}}

\newcommand{\bham}{\affiliation{School of Physics and Astronomy \& Institute for Gravitational Wave Astronomy, University of Birmingham, \\ Birmingham, B15 2TT, United Kingdom}}

\newcommand{\milan}{\affiliation{Dipartimento di Fisica ``G. Occhialini'', Universit\'a degli Studi di Milano-Bicocca, Piazza della Scienza 3, 20126 Milano, Italy}}
\newcommand{\infn}{\affiliation{INFN, Sezione di Milano-Bicocca, Piazza della Scienza 3, 20126 Milano, Italy}}

\begin{document}
\title{
Characterization of merging black holes with two precessing spins
}

\author{Viola De Renzis$\,$\orcidlink{0000-0001-7038-735X}}
\email{v.derenzis@campus.unimib.it}
\milan \infn

\author{Davide Gerosa$\,$\orcidlink{0000-0002-0933-3579}}

\milan \infn \bham

\author{Geraint Pratten$\,$\orcidlink{0000-0003-4984-0775}}
 \bham

\author{Patricia Schmidt$\,$\orcidlink{0000-0003-1542-1791}
}
 \bham

 \author{Matthew Mould$\,$\orcidlink{0000-0001-5460-2910}}
  \bham


\date{\today}

\begin{abstract}

Spin precession in merging black-hole binaries is a treasure trove for both astrophysics and fundamental physics. %
There are now well-established strategies to infer from gravitational-wave data whether at least one of the two black holes is precessing. In this paper we tackle the next-in-line target, namely the statistical assessment that the observed system has two precessing spins. We find that the recently developed generalization of the effective precession spin parameter $\chi_\mathrm{p}$ 
is a well-suited estimator to this task. With this estimator, the occurrence of two precessing spins is a necessary (though not sufficient) condition to obtain values $1<\chi_\mathrm{p}\leq 2$. %
Confident measurements of gravitational-wave sources with $\chi_\mathrm{p}$ values in this range %
can be taken as a conservative assessment that the binary presents two precessing spins. We investigate this argument using a large set of >100 software injections assuming anticipated LIGO/Virgo sensitivities for the upcoming
fourth
observing run,
O4.
Our results are very encouraging,
suggesting that, if such binaries exist in nature and merge at a sufficient rate, current interferometers 
are likely to deliver the first confident detection of merging black holes with two precessing spins.
We investigate prior effects and waveform systematics and, though these need to be better investigated,
did not find any confident false-positive
case among all the configurations we tested. 
Our assessment should thus be taken as conservative.

\end{abstract}

\maketitle

\section{Introduction}
Black-hole (BH) binary spin precession is a key feature of the relativistic two-body problem \cite{1994PhRvD..49.6274A,1995PhRvD..52..821K}.  %
Spin-spin and spin-orbit couplings in general relativity
cause the orbital angular momentum \textbf{L}  and the BH spins  ${\bf S}_{1,2}$ to jointly precess about the direction of the total angular momentum $\bm{J}=\bm{L}+{\bf S}_{1}+{\bf S}_{2}$.
This motion induces modulations to both the amplitude and the phase of the emitted gravitational waves (GWs).

Measurements of spin precession have important repercussions in both astrophysics and fundamental physics. For the stellar-mass BH binaries observed by LIGO and Virgo \cite{2015CQGra..32g4001L,2015CQGra..32b4001A}, %
spin precession provides
unique leverage to discriminate between BH binaries formed in isolation and those assembled in dynamically in stellar clusters~\cite{2013PhRvD..87j4028G,2016ApJ...832L...2R,2017CQGra..34cLT01V,2017MNRAS.471.2801S}.
For the supermassive BH binaries targeted by LISA~\cite{2017arXiv170200786A}, spin measurements will provide information on, e.g., the occurrence of prolonged phases of disk accretion~\cite{2014ApJ...794..104S,2021MNRAS.501.2531S}.
GW observations of precessing binary BHs also allow us to constrain modified theories of gravity, especially those with parity-violating interactions caused by additional fields~\cite{2018PhRvD..98f4020L}.

While the masses of LIGO/Virgo events are usually well measured, %
spin effects provide a subdominant contribution to the emitted
radiation
and are thus considerably more challenging to characterize.  %
At present, an unambiguous measurement of BH-binary spin precession is one of the holy grails of observational GW astronomy.

{Data from the first three observing runs of the LIGO/Virgo network
have provided some evidence for individual BH mergers with highly
precessing spins~\cite{2019PhRvX...9c1040A,2021PhRvX..11b1053A,2021arXiv210801045T,2021arXiv211103606T}. The most suggestive indication is that from
GW200129\_065458, where Refs.~\cite{2021arXiv211211300H,2022PhRvL.128s1102V} found strong evidence for
BH-binary spin precession, while Ref.~\cite{2022arXiv220611932P} raised potential issues in
the glitch mitigation analysis.} %
For the case of GW190521, a strong precession signature was also reported \cite{2021PhRvX..11b1053A}, though potential degeneracies with the eccentricity still need to be fully understood \cite{2020ApJ...903L...5R}. %
Collective evidence for spin precession was reported in the context of  BH binary populations, with all current fits requiring some misaligned spins at high confidence~\cite{2021arXiv211103634T,2022ApJ...937L..13C,2022arXiv220512329M}.

Upcoming instrumental upgrades to the LIGO/Virgo (and hopefully KAGRA) network \cite{2020LRR....23....3A} are posed to provide increasingly accurate spin measurements. %
{It is therefore not unreasonable to predict that the next observing
run will deliver a confident, unambiguous identification of BH-binary
spin precession.} %
Crucially, measuring orbital-plane precession corresponds to inferring that {at least one of the two BHs} has a misaligned spin~\cite{1994PhRvD..49.6274A}. Inferring the presence of two misaligned spins requires extracting
even feebler signatures from the signal,
which are related to spin-spin (as opposed to spin-orbit) terms in the BH binary equations of motion.

This paper tackles such a next-in-line target. We perform $>100$ software injections with realistic LIGO/Virgo sensitivity
and demonstrate that signals with large-but-not-extreme signal-to-noise ratio (SNR) $\gtrsim 20$ allow us to detect two-spin effects \emph{already in the next LIGO/Virgo observing run} (O4). Of course, this statement relies on the assumption that merging binaries with two large precessing spins exist and can merge efficiently. But if such GW sources
are out there
in the Universe, the next LIGO/Virgo run might provide the first observational constraints of their properties.

Compared to previous analyses which include two precessing spins (e.g.,~\cite{2019PhRvD.100b4059K,2021PhRvD.104j3018B,2022PhRvD.105b4045V})
our investigation relies on a state-of-the-art reformulation of the precession estimator $\chi_\mathrm{p}$~\cite{2021PhRvD.103f4067G}. 
This generalizes the commonly used expression~\cite{2015PhRvD..91b4043S} by employing a rigorous post-Newtonian(PN) average over the joint evolution of
\emph{both}
spins.
Measurements of such an augmented $\chi_\mathrm{p}$ for current GW events have been presented in Refs.~\cite{2021PhRvD.103f4067G,2022CQGra..39l5003H}. Crucially for this paper, the precession-averaged estimator presents an exclusion region $1< \chi_\mathrm{p}\leq 2$ that can
\emph{only}
be populated by binaries with two precessing spins. Measuring a binary with $\chi_\mathrm{p}>1$ at some large confidence (GW astronomers often use the $90\%$ Bayesian credible interval), would allow us to claim the first detection of binary BH physics sourced by two precessing spins.

Our paper is organized as follows. In Sec.~\ref{sec:methods} we present our methodology, including details on the adopted precession estimator as well as the implemented parameter-estimation pipeline.
In Sec.~\ref{sec:result}  we present the results of our analysis. In particular, we characterize
(i)~the SNR dependence on the resulting $\chi_{\rm p}$ posterior distributions, (ii)~the statistical behavior of large ensembles of sources, (iii)~the impact of the prior, and (iv)~the relevance of waveform systematics. Our conclusions are reported in Sec.~\ref{sec:conclusions}.
In the following we employ geometric units $G=c=1$.

\section{Methods}
\label{sec:methods}
\subsection{Spin precession estimators}
\label{subsec:SPestimators}

While the full BH-binary spin properties are in principle described by six degrees of freedom (three components for two spin vectors), a considerable amount of effort has been devoted to identifying a reduced number of parameters that encapsulate most of the information. These are often derived in a PN framework, %
with the most widely used quantities being the effective aligned spin $\chi_\mathrm{eff}$ \cite{2001PhRvD..64l4013D,2008PhRvD..78d4021R,2015PhRvD..92f4016G} and the effective precessing spin $\chi_{\rm p}$ \cite{2021PhRvD.103f4067G,2015PhRvD..91b4043S}.
Alternative approaches include extending the precession estimator to a two-dimensional vector \cite{2021PhRvD.103h3022T}, exploiting the precession/nutation amplitudes and frequencies \cite{2021PhRvD.103l4026G,2022PhRvD.106b4019G}, and computing the fraction of the SNR contained in the spin modulations \cite{2020PhRvD.102d1302F}. %

Let us consider a BH binary where
$q=m_2/m_1\leq1$ is the mass ratio, $\chi_{i}%
\in [0,1]$ are the dimensionless spin magnitudes, $\theta_i$ are the angles between the spins and the orbital angular momentum, and $\Delta\Phi$ %
is the angle between the projections of the two spins onto the orbital plane.

The effective aligned spin is defined as \cite{2001PhRvD..64l4013D}
\begin{equation}
\chi_\mathrm{eff}=\frac{{\chi_{1} \cos\theta_1}+ q{\chi_{2} \cos\theta_2}}{1+q}\,.
\label{eq:chieffective}
\end{equation}
This is the spin quantity that affects the GW phase at lowest order and is a constant of motion at 2PN \cite{2008PhRvD..78d4021R,2015PhRvD..92f4016G}. The effective spin $\chi_\mathrm{eff}$  was recognized as the best measured spin parameter since the very first GW detections, the key reason being that it directly impacts the length of the signal.

The spin-precession parameter $\chi_\mathrm{p}\propto |d \hat{\bm L}/dt|$ tracks the change of the direction of the orbital angular momentum ${\bm L}$
over time $t$
~\cite{2015PhRvD..91b4043S, 2021PhRvD.103f4067G}. It was originally introduced by \citeauthor{2015PhRvD..91b4043S}~\cite{2015PhRvD..91b4043S} as a building block toward the construction of precessing waveforms.
Their definition reads \begin{equation}
\chi_\mathrm{p}^{\rm (heu)} =  \text{max}\left(\chi_{1}\sin\theta_{1},q \frac{4q+3}{4+3q}\chi_{2}\sin\theta_{2}\right)\,,
\label{eq:chip_heuristic}
\end{equation}
 which in this paper we refer to as ``{heuristic} $\chi_\mathrm{p}$.''
This precessing spin parameter is defined in the domain $\chi_\mathrm{p}^{(\mathrm{heu})}\in[0,1]$. Unlike $\chi_{\rm eff}$, the parameter  $\chi_\mathrm{p}$ depends on the projections of the spins onto the orbital plane, $\chi_i\sin\theta_i$, implying that a confident measurement of $\chi_{\rm p}>0$ requires that at least one of the two BH spins was misaligned before merger, and hence that the system was precessing. %

\citeauthor{2021PhRvD.103f4067G}~\cite{2021PhRvD.103f4067G} recently pointed out that Eq.~(\ref{eq:chip_heuristic}) was derived by preferentially selecting some terms when averaging over the spin motion. Mathematically, this is reflected in the maximization operation reported in Eq.~(\ref{eq:chip_heuristic}), which selects one of the two BHs as dominant to the precession dynamics, thus obfuscating two-spin effects. Relaxing this approximation yields a generalized parameter \cite{2021PhRvD.103f4067G}
\begin{equation}
\begin{split}
\chi_\mathrm{p}^{(\mathrm{gen})}=&\Bigg[\bigg(\chi_{1}\sin\theta_{1}\bigg)^{2}+\bigg(q \frac{4q+3}{4+3q}\chi_{2}\sin\theta_{2}\bigg)^{2}\\
&+2q \frac{4q+3}{4+3q}\chi_{1}\chi_{2}\sin\theta_{1}\sin\theta_{2}\cos\Delta\Phi\Bigg]^{\frac{1}{2}}\,,
\label{eq:chip_generalized}
\end{split}
\end{equation}
where the angles  $\theta_1(t)$, $\theta_2(t)$, and $\Delta\Phi(t)$ all vary jointly with time. This can be averaged over a single precession cycle to obtain %
\begin{equation}
\chi_\mathrm{p}^{(\mathrm{av})}= \frac{1}{\tau} { \int_0^{\tau} \chi_\mathrm{p}^{(\mathrm{gen})}(t) dt },
\label{eq:chip_averaged}
\end{equation}
where $\tau$ is the precession period. We argue this should be regarded as a more solid estimator because, although it is not a constant of motion like $\chi_{\rm eff}$,
it at most
varies only over the longer radiation-reaction timescale. In the following, we refer to Eq.~(\ref{eq:chip_averaged}) as the ``averaged $\chi_{\rm p}$'' parameter.
 In practice, we perform the integral in Eq.~(\ref{eq:chip_averaged}) using a 2PN quasiadiabatic approach where the precession cycle is parametrized by $S(t) =  | {\bf S}_1(t)+{\bf S}_2(t)|$ \cite{2015PhRvL.114h1103K,2015PhRvD..92f4016G}. We refer the reader to Ref.~\cite{2015PhRvD..92f4016G} for details on the derivation of Eqs.~(\ref{eq:chip_generalized}) and (\ref{eq:chip_averaged}), but stress that the starting point is simply the derivative $d \hat{\bm L}/dt$.

The reformulation of the precession parameter defines an extended range
$\chi_\mathrm{p}^{(\mathrm{av})}\in [0,2]$. As shown in Ref.~\cite{2021PhRvD.103f4067G},  the heuristic and averaged definitions of $\chi_{\rm p}$ have the same single-spin limit, which implies that the range of the latter cannot be freely absorbed with a normalization factor. From Eq.~(\ref{eq:chip_generalized}), it is immediate to see that $\chi_{\rm p}>1$ requires both $\chi_1\sin\theta_1\neq 0$ and $\chi_2\sin\theta_2\neq 0$, i.e.,
the binary must have two precessing spins.
Such sources \emph{can} lie in $0\leq\chi_\mathrm{p}\leq1$, but both spins being misaligned is requisite in the two-spin domain.
From Eq.~(\ref{eq:chip_generalized}), there is a larger volume of parameter space where $\chi_{\rm p}>1$ for comparable mass binaries $q\lesssim 1$ compared to asymmetric sources with $q\ll 1$. %
This is expected, as two-spin effects are highly suppressed in the in the low-mass ratio limit where $S_2/S_1 \propto q^2\ll 1$ (cf. Ref.~\cite{2020PhRvR...2d3096P} for more work on spin precession in asymmetric binaries).

\subsection{Parameter estimation pipeline}
\label{subsec:PE}

As is common practice in GW parameter estimation, we employ the following fifteen parameters to describe compact-binary coalescences:
detector-frame total mass $M=m_1+m_2$, mass ratio $q=m_2/m_1$, dimensionless spin magnitudes  $\chi_{1,2}$, tilt angles $\theta_{1,2}$, azimuthal spin angle $\Delta\Phi$, azimuthal angle $\phi_{JL}$ between the total and orbital angular momenta,
luminosity distance $D_{L}$, right ascension  $\alpha$, declination $\delta$, polar angle $\theta_{JN}$ between total angular momentum and the line of sight, polarization $\psi$, time  $t_{c}$ and phase  {$\phi_{c}$} of coalescence \cite{2015PhRvD..91d2003V}.  %

We explore the joint Bayesian posterior distribution of these parameters under a Gaussian noise likelihood (e.g.,~\cite{2008PhRvD..78b2001V}) using the parallelized {\sc Bilby} pipeline \cite{2019ApJS..241...27A,2020MNRAS.498.4492S} and its underlying {\sc Dynesty} implementation of nested sampling \cite{2020MNRAS.493.3132S}. Our runs make use of 2048 live points, a number of autocorrelation equal to 50
a random walk sampling method, and a likelihood marginalized over time and distance. Runs are halted when the log-evidence gain falls below 0.1. %

We consider a three-detector network consisting of LIGO Livingston, LIGO Handford, and Virgo with their
projected sensitivities
for the upcoming fourth observing run O4~\cite{2020LRR....23....3A}. %
We consider data segments of $4$ s, set a lower frequency cutoff of $20$ Hz, assume a sampling frequency of $2048$ Hz, and zero noise. %
Time-varying quantities are quoted when the detector-frame emission frequency of the dominant mode is $20$ Hz. This choice has a negligible impact on the averaged $\chi_{\rm p}$  estimator  because it only varies on the long inspiral timescale of the binary evolution; see Ref.~\cite{2021PhRvD.103f4067G}.
Unless stated otherwise, we quote our results using medians and 90\% equal-tailed credible intervals.

We adopt uninformative priors as commonly used in current LIGO/Virgo analyses~\cite{2019PhRvX...9c1040A,2021PhRvX..11b1053A,2021arXiv210801045T,2021arXiv211103606T}.
Specifically, priors on the
masses are chosen to be uniform in
$m_{1,2}\in[5,100]M_{\odot}$, with further constraints imposed on the mass ratio $q\in[1/8,1]$ and detector-frame chirp mass ${M_{\rm c}}\in[10,60]M_{\odot}$. %
For most of our runs, priors on the spins are taken to be unform in magnitude  $\chi_{1,2}\in[0,0.99]$ and isotropic in directions. In the following, we will refer to this as our ``standard'' spin prior. To better explore prior effects,
some of our runs are performed with a ``volumetric'' spin prior  $p(\chi_i)\propto \chi_{i}^2$ , corresponding to spin vectors that are uniformly drawn in volume (e.g.,~\cite{2017PhRvL.119y1103V,2018PhRvD..98d4028C,2019PhRvX...9c1040A}).  The luminosity distance prior is taken to be uniform in comoving volume with $D_L\in [100, 5000]$ Mpc. %

\begin{figure}
    \includegraphics[width=0.95\columnwidth]{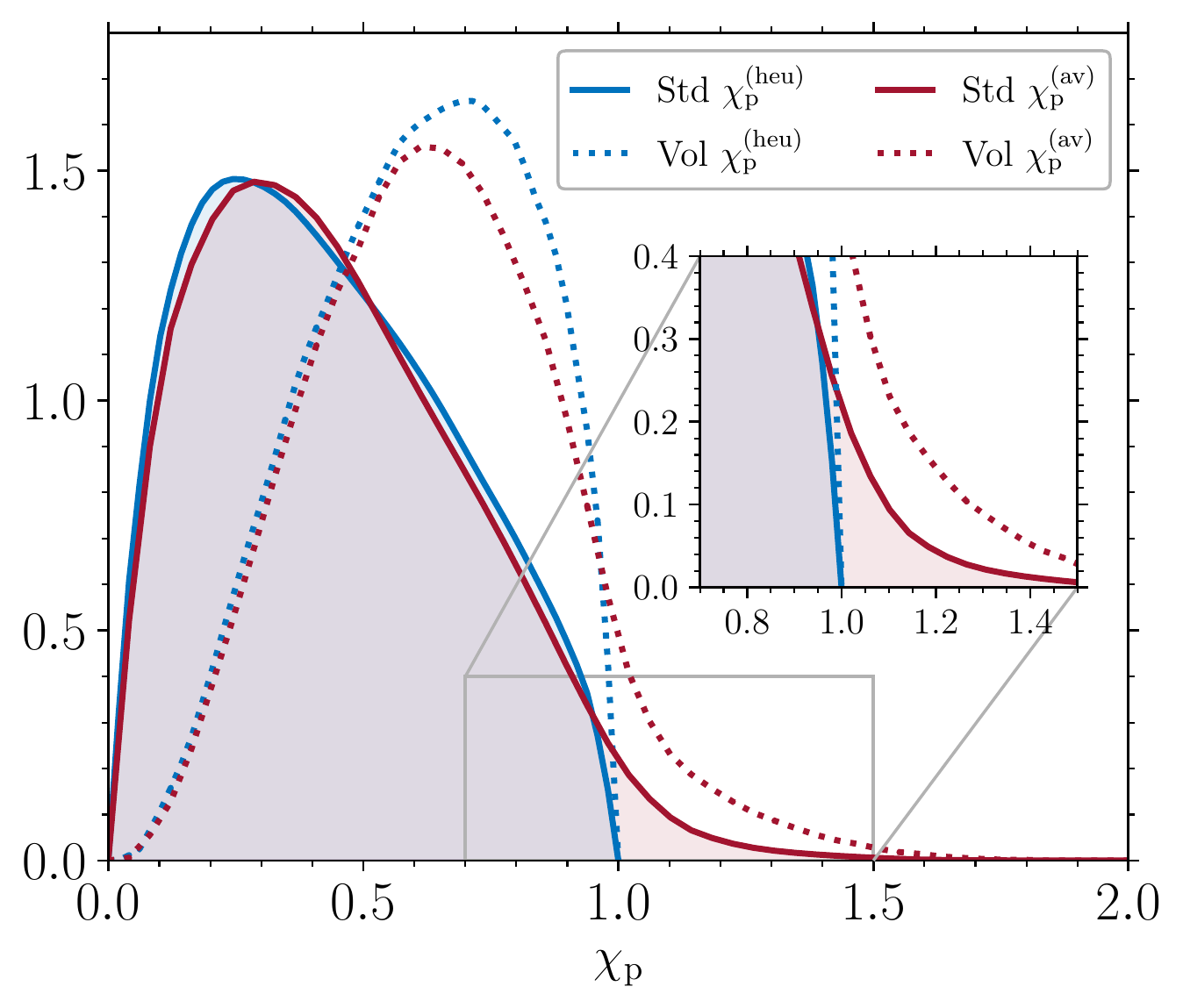}
    \caption{Prior distributions for the heuristic (blue) and averaged (red)  $\chi_\mathrm{p}$  estimators.  Dotted curves are computed assuming the standard spin prior $p(\chi)={\rm const.}$; dashed curves instead assume a volumetric prior  $p(\chi)\propto \chi^2$.  %
}   \label{fig:prior}
\end{figure}

Figure~\ref{fig:prior} shows the resulting prior probability density for the heuristic and the averaged $\chi_\mathrm{p}$ definitions. %
At low values of $\chi_\mathrm{p}$, the prior distributions of the two estimators %
are qualitatively very similar.
This behavior was explicitly imposed in Ref.~\cite{2021PhRvD.103f4067G} when generalizing the $\chi_{\rm p}$ definition. By construction, the region of $1<\chi_\mathrm{p} \leq 2$ is not allowed for the heuristic formulation and, consequently, the prior distribution of $\chi_\mathrm{p}^{(\mathrm{heu})}$ is steeply truncated at $\chi_\mathrm{p}=1$. On the other hand, the prior distribution of $\chi_\mathrm{p}^{\rm (av)}$ extends into the two-spin region $\chi_\mathrm{p}>1$. However, under these commonly used assumptions, the tail at large $\chi_{\rm p}$ values is very sparsely populated. From Eq.~(\ref{eq:chip_generalized}), reaching
$\chi_{\rm p}\approx2$
requires
systems with
$q\approx1$,
$\chi_{1,2}\approx1$,
$\theta_{1,2}\approx\pi/2$,
and
$\Delta\Phi\approx0$.
Such a strong prior suppression is a key element of our analysis and suggests that current GW data are being analyzed with a prior that strongly disfavors the region of
parameter space that is exclusive to two-spin physics. %
Although still present, this effects is less prominent for the volumetric spin prior. More quantitatively, we find $p(\chi_{\rm p}>1) =$ 0.02 and 0.07 for the averaged estimator under the standard and volumetric prior, respectively. %

\begin{figure*}\centering
    \includegraphics[width=0.9\textwidth]{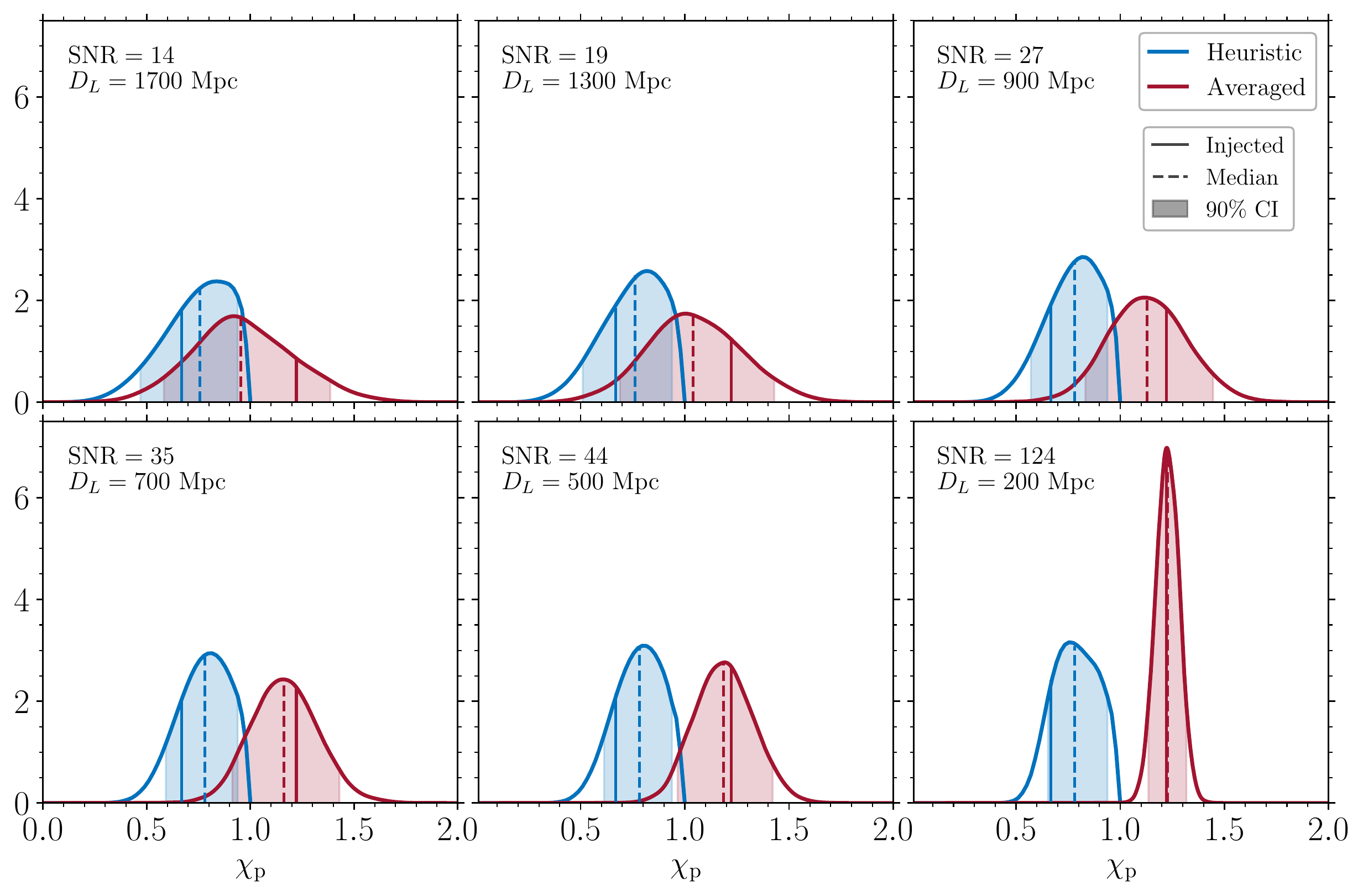}
    \caption{Posterior distribution of the heuristic (blue) and averaged (red)  $\chi_\mathrm{p}$ precession estimator for the single-system series described in Sec.~\ref{sub:differentSNR}. Panels from left to right and top to bottom shows results for the same source injected at increasing SNRs and decreasing luminosity distance $D_L$. Solid and dashed lines indicate the true value and the median of the recovered posterior. The shaded areas indicate the $90\%$ (CI) credible intervals.   %
}   \label{fig:seriesDL}
\end{figure*}
\begin{figure}
    \includegraphics[width=\columnwidth]{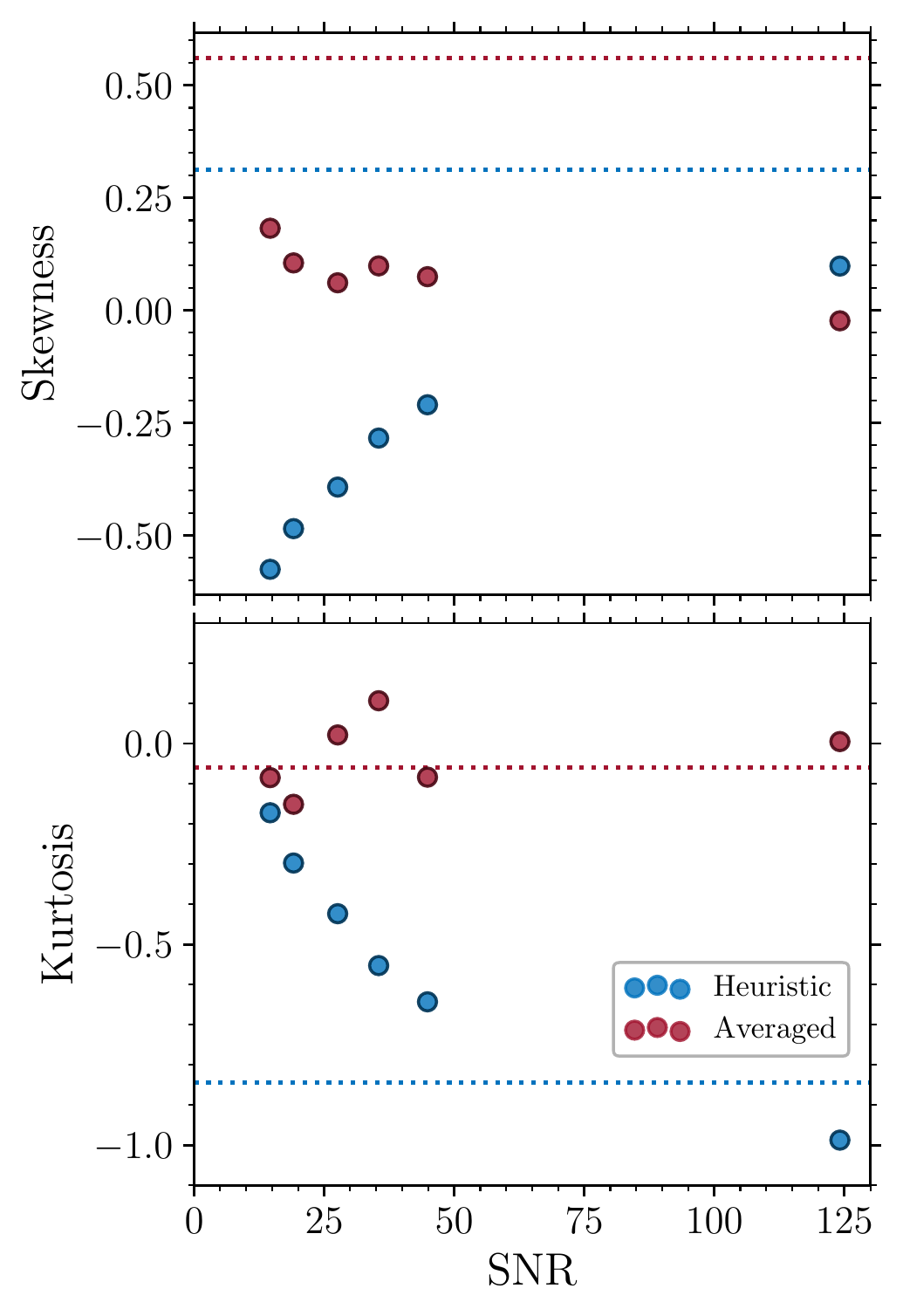}
    \caption{Skewness (upper panel) and kurtosis (lower panel) of the $\chi_\mathrm{p}$ posterior distribution as a function of the SNR for the six injections described in Sec.~\ref{sub:differentSNR}.  Blue and red scatter points refer to the heuristic and averaged $\chi_{\rm p}$ definition, respectively. The dotted lines represent the values of the skewness and kurtosis  calculated from the priors.    %
 }
 \label{fig:skew}
\end{figure}

For the majority of our runs, we employ the {\sc IMRPhenomXPHM}~\cite{2020PhRvD.102f4001P,2020PhRvD.102f4002G,2021PhRvD.103j4056P} waveform model for both  injection and recovery. This is a state-of-the-art frequency-domain approximant that captures spin precession without relying on a single-spin approximation.  Selecting the same model for both injection and recovery allows us to first isolate statistical effects without systematics. Waveform systematics are then explored with a dedicated analysis  where we select different models for injection and recovery. In particular, we use the time-domain model {\sc IMRPhenomTPHM}~\cite{2022PhRvD.105h4040E} as well as the numerical-relativity
surrogate
{\sc NRSur7dq4}~\cite{2019PhRvR...1c3015V}.
When recovering with {\sc NRSur7dq4}, we restrict our prior to $q>1/6$, which corresponds to the extended range of validity of the model.
Because this approximant only covers $\ssim 20$ orbits before merger,
we also restrict our priors to $m_{1,2} \in [35,150]M_{\odot}$ and $M_{\rm c} \in [40,60]M_{\odot}$ to ensure the signal is fully above the low-frequency cutoff  of $20$ Hz.

\section{Results}
\label{sec:result}

\subsection{Single-system series}
\label{sub:differentSNR}

As a first step, we highlight the main implications of the
$\chi_\mathrm{p}$ reformulation
on GW parameter estimation. %
To this end, we present a series of six software injections where the same binary is observed at different SNRs. We select a source with $\chi_\mathrm{p}^{(\mathrm{heu})}= 0.67$ and $\chi_\mathrm{p}^{(\mathrm{av})}=1.22>1$, which thus contains two prominently precessing spins. 
In particular, the injected system has %
$M=54.1\,M_\odot$, $q=0.96$, $\chi_{1}= 0.56$, $\chi_{2}= 0.7$, $\theta_{1,2}=\pi/2$, $\Delta\Phi=0.1$, $\theta_{JN}= 1.0$, $\phi_{JL}= 1.0$,
$\alpha=0.75$, $\delta=0.5$, $\psi= 1.0$, $\phi_c= \pi/4$, $t_{c}=0.0$.  We select increasing values of the  luminosity distance $D_{L}=$ 200, 500, 700, 900, 1300, 1700 Mpc while keeping the detector-frame %
mass $M$ fixed. The corresponding three-detector network SNRs are $\rho=$ 124, 44, 35, 27, 19, and 14. We use  the {\sc IMRPhenomXPHM} waveform model for both injection and recovery and employ standard uninformative priors.

Our results are illustrated in Fig.~\ref{fig:seriesDL}.
As one moves from the lowest to the highest value of the SNR, the recovered posteriors of both the averaged and the heuristic $\chi_\mathrm{p}$ converge to the injected values. %
This is expected because we have used the same signal model for injection and recovery and we are not considering a specific noise realization.

For the system with the lowest SNR $\rho = 14$, the posteriors of our two $\chi_\mathrm{p}$
definitions
largely overlap. Quoting median and 90\% credible interval, we find $\chi_\mathrm{p}^{(\mathrm{av})}=0.95^{+0.43}_{-0.37}$, which implies that we cannot confidently tell that the source has two misaligned spins. %
As the SNR increases, so does our ability to infer that the binary has two precessing spins.
For the system with the largest SNR $\rho=124$, the two marginalized $\chi_\mathrm{p}$ distributions are almost completely detached. %
The posterior of the heuristic $\chi_\mathrm{p}$ is by definition truncated at $\chi_\mathrm{p}=1$ because Eq.~(\ref{eq:chip_heuristic}) allows only for the contribution from a single, dominant spin.
On the contrary, considering our averaged definition yields $\chi_\mathrm{p}=1.22^{+0.09}_{-0.09}$  for $\rho=124$, implying one infers the presence of two precessing spins with a credibility of   $p(\chi_\mathrm{p}>1) = 99.9\%$.
 
Figure~\ref{fig:seriesDL} also shows that the posterior of the averaged $\chi_\mathrm{p}$ is closer to a Gaussian compared to that of the heuristic estimator.
This indicates that, if a significant non-Gaussianity in the heuristic  $\chi_\mathrm{p}$ posterior were to
appear
in GW data,
it could be taken as a potential indication that
 some additional two-spin physics is present but is being missed because of the suboptimality of the employed estimator.
 
This argument is further explored in Fig.~\ref{fig:skew}, where we show the skewness and kurtosis  for the same six injections  presented in Fig.~\ref{fig:seriesDL}. These quantities are related to the third and fourth moments of the distribution and describe the departure from Gaussianity; both are zero for normally distributed data, with the skewness quantifying the left-right asymmetry and the kurtosis quantifying the weight of the tails~\cite{2019sdmm.book.....I}.  Figure~\ref{fig:skew} shows that both skewness and kurtosis of the averaged $\chi_\mathrm{p}$ are approximately distributed around 0. %
On the other hand, the skewness (kurtosis) of the heuristic $\chi_\mathrm{p}$ strongly increases (decreases) with the SNR. This indicate that (i) the heuristic $\chi_\mathrm{p}$ posteriors have a thinner tails compared a normal distribution and that (ii) their left tail is more pronounced compared to the right tail. These features can be taken as a quantification of the artificial cutoff at $\chi_{\rm p}=1$, an assumption that is naturally relaxed when considering the averaged $\chi_{\rm p}$ estimator.

\subsection{Parameter-space exploration}
\label{sub:uniformchip}

Using the same settings, we now target the statistical properties emerging from a large number of injected signals. Ideally, one would  want to inject signals drawn from the prior (this is necessary, for instance, to present a probability-probability plot \cite{chambers1983graphical}). In our case, such a procedure would be highly suboptimal and ultimately computationally intractable because, as shown in Fig.~\ref{fig:prior}, the two-spin region with $\chi_\mathrm{p}>1$ corresponds to a very low prior volume (where from now on in the paper we only refer to the average formulation of $\chi_\mathrm{p}$). Most of the injections would thus be placed in the region where only one of the two spins dominates. We thus opt for
an injection distribution with $\chi_\mathrm{p}$ uniform in $[0,2]$
which, although of dubious astrophysical relevance, is well suited to assess the statistical property of the proposed estimator.
More precisely, we draw values of $\chi_\mathrm{p}$ and then reweight samples of the intrinsic binary properties drawn from the
uninformative
prior
(Sec.~\ref{subsec:PE})
to the injection distribution using an acceptance/rejection scheme with an absolute numerical tolerance of 0.04 between the original and resampled values of $\chi_\mathrm{p}$.
We have verified that this choice does not significantly impact our results. Since precession effects are subdominant in the waveform,
when selecting the extrinsic properties for injections
we only consider sources with $\rho>20$, i.e., $\approx2$ times larger than the current detection threshold~\cite{2021arXiv211103606T}. %

\begin{figure*}
    \includegraphics[width=0.77\textwidth]{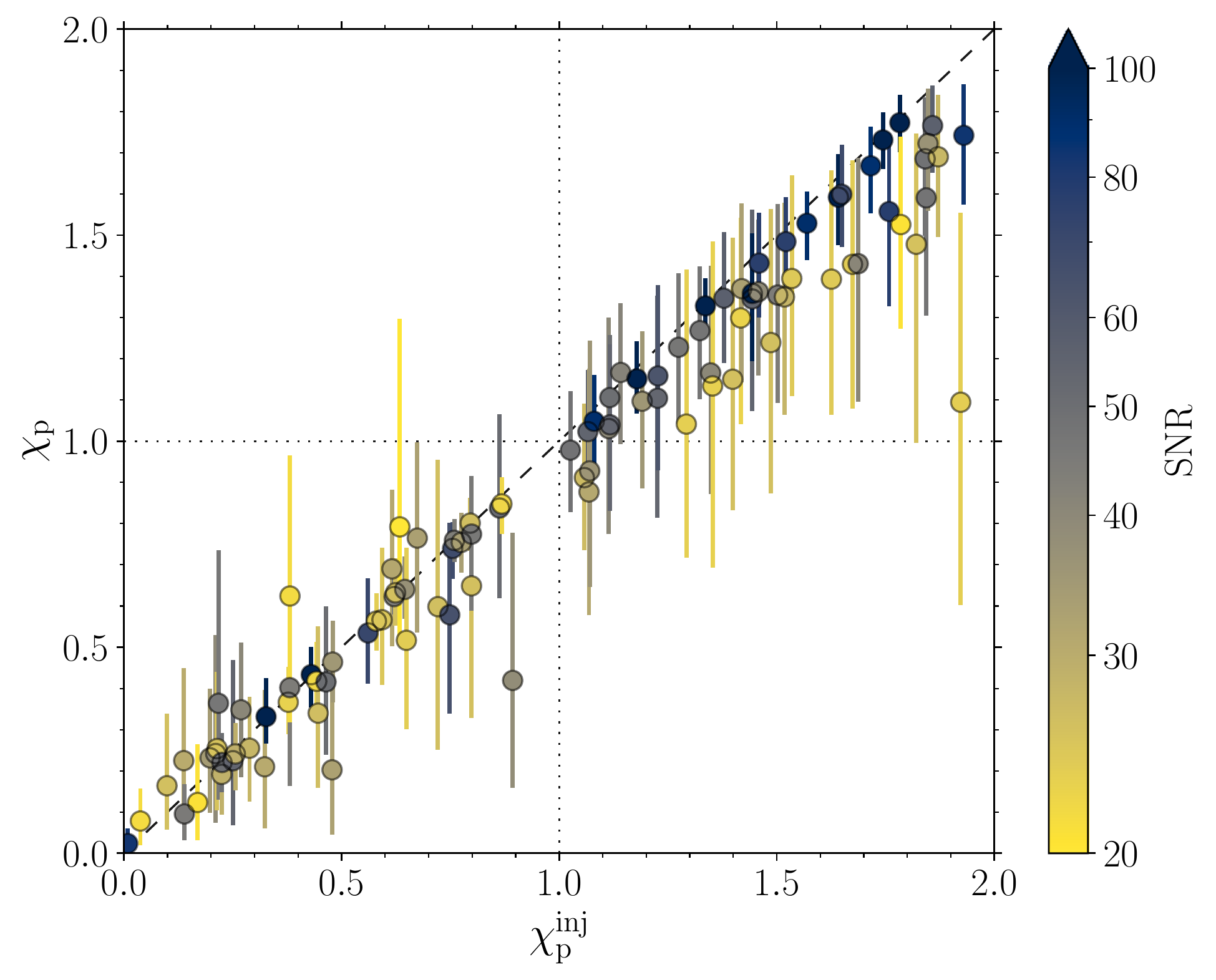}
    \caption{Set of 100 injections obtained by reweighting the averaged $\chi_\mathrm{p}$ prior toward a uniform distribution in $[0,2]$.
The medians (scatter points) and symmetric 90\% credible intervals (error bars) of the recovered posteriors are plotted against the true values $\chi_{\rm p}^{\rm inj}$. Vertical and horizontal dotted lines indicate $\chi_\mathrm{p}=1$ while the dashed diagonal line corresponds to $\chi_\mathrm{p}=\chi_\mathrm{p}^\mathrm{inj}$, i.e., successful recovery. The three-detector SNRs of the injected sources are reported on the color scale.%
}   \label{fig:55_injections}
\end{figure*}

\begin{figure}
    \includegraphics[width=\columnwidth]{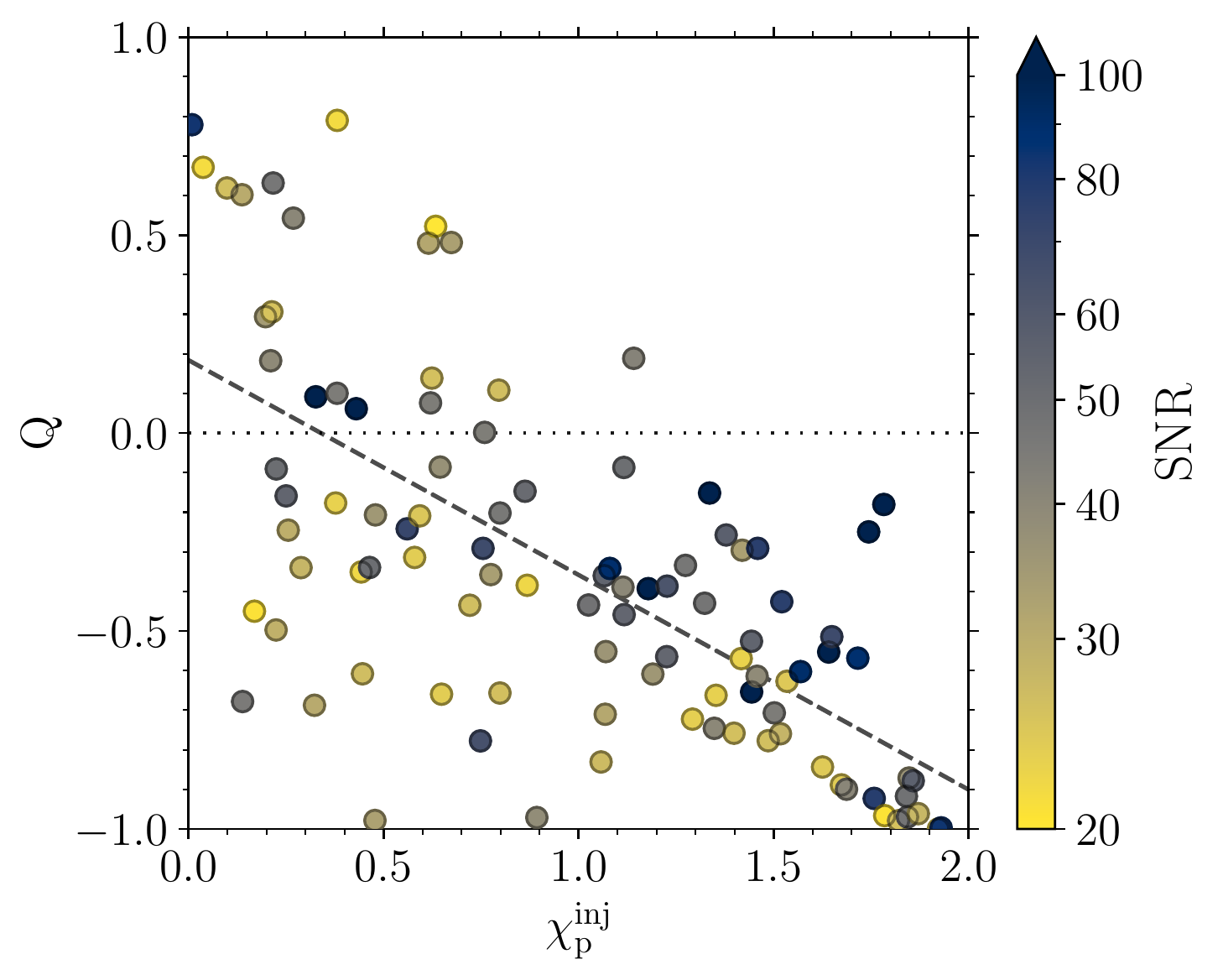}
    \caption{Adjusted
posterior quantile $Q$ for the posterior distribution of the averaged $\chi_{\rm p}$ parameter. Sources above (below) the horizontal dashed %
line indicate cases cases where $\chi_{\rm p}$ is overestimated (underestimated). To guide the eye, the diagonal dashed line shows a simple linear fit $Q= -0.54\chi_{\rm p}^{\rm inj} + 0.18$. The color scale indicates the SNRs of the sources.
}   \label{fig:quantiles}
\end{figure}

\begin{figure*}
    \includegraphics[width=0.85\textwidth]{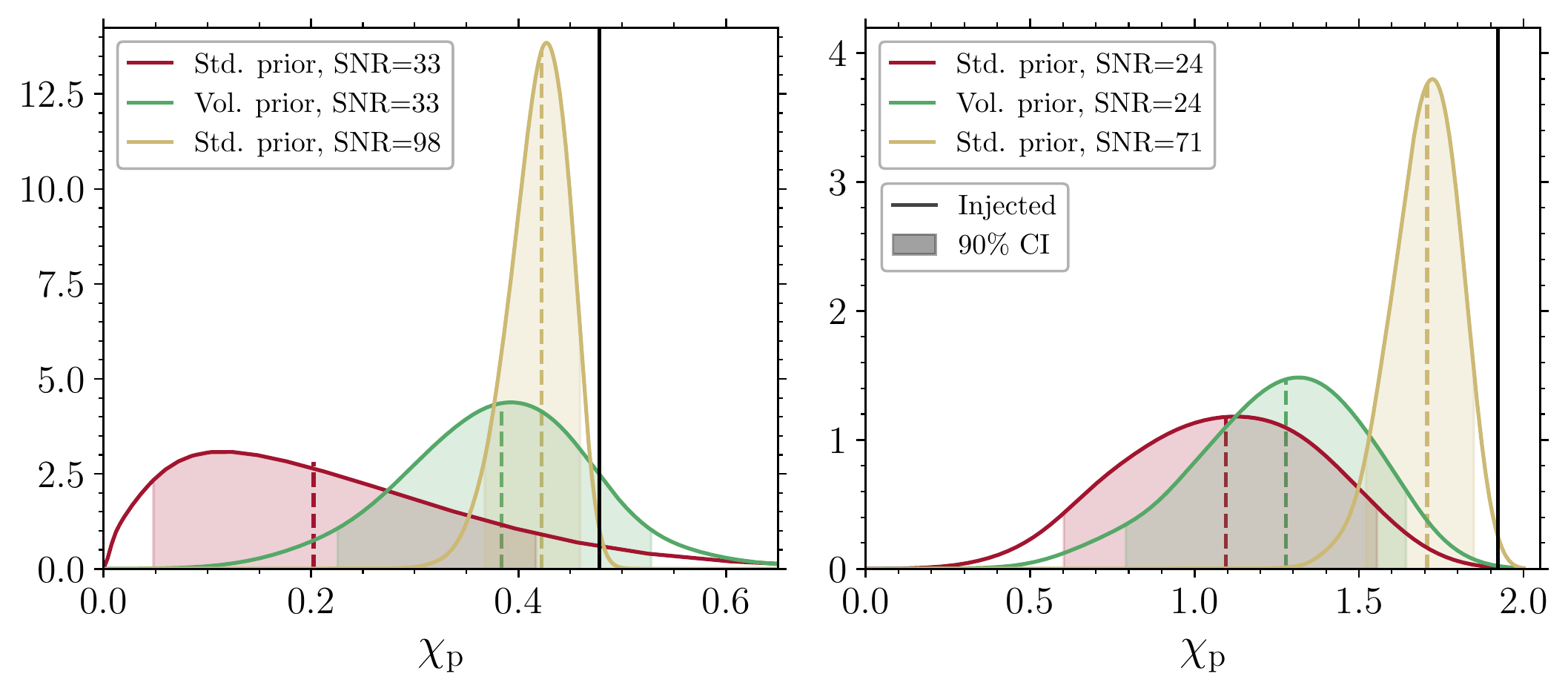}
    \caption{Recovery  of the averaged $\chi_{\rm p}$ estimator with different priors and SNRs. The left and right panel shows results for sources injected with $\chi_\mathrm{p}=0.48$ and $\chi_\mathrm{p}=1.92$, %
respectively (black vertical lines). The red and green posterior distributions are   obtained under the standard and volumetric priors, respectively, and the same SNRs used in Sec.~\ref{sub:uniformchip}. For the  yellow distributions, the SNR was boosted by a factor %
$\approx3$.
 Dashed lines indicate the medians of the posteriors %
 while the shaded area indicates  the 90\% credible interval. %
}   \label{fig:outliers}
\end{figure*}

Figure~\ref{fig:55_injections} shows the recovered posteriors of the averaged $\chi_\mathrm{p}$ parameter as a function of the
true values for 100 such injections.
One can divide the parameter space into four distinct regions, acting much like a confusion matrix in statistics.
\begin{enumerate}[label=(\roman*)]
\item {\it True negatives} (bottom-left quadrant in Fig.~\ref{fig:55_injections}): injected $\chi_\mathrm{p}<1$ and recovered  $\chi_\mathrm{p}<1$. The injected configurations are not unique to sources with two precessing spins and are recovered as such.
\item {\it False positives} (top-left quadrant in Fig.~\ref{fig:55_injections}): injected $\chi_\mathrm{p}<1$ and recovered  $\chi_\mathrm{p}>1$. For these sources, one infers the presence two precessing spins even if they might not be present.  
\item {\it False negatives} (bottom-right quadrant in Fig.~\ref{fig:55_injections}): injected $\chi_\mathrm{p}>1$ and recovered  $\chi_\mathrm{p}<1$. In this region  sources have two precessing spins but one is not able infer their occurrence from the
signal. 
\item {\it True positives} (top-right quadrant in Fig.~\ref{fig:55_injections}): injected $\chi_\mathrm{p}>1$ and recovered  $\chi_\mathrm{p}>1$. These sources are characterized by two precessing spins and one can successfully infers that this is the case.
 \end{enumerate}

For each posterior distribution, we compute the fraction of the samples in each of these four regions and then compute the arithmetic mean over the injected sample (this is equivalent to assuming a flat population prior on $\chi_{\rm p}$ because our injections are distributed uniformly). We report 47.45\% of true negatives, 0.55\% of false positives, 7.01\% of false negatives, and 44.98\% of true positives. %

From Fig.~\ref{fig:55_injections}, the signals with higher SNR lie closer to the injected values and present thinner posterior distributions, as expected.
In the true negative region, %
the recovered posteriors are distributed around the true value without evident systematic trends. %
 On the other hand, when $\chi_\mathrm{p}>1$, the recovered posteriors systematically underestimate the true value. While this is, in general, true for most systems, in a few cases this is sufficient to cause false negatives. %

One can further quantify this behavior using the 
adjusted
posterior quantile
\begin{equation}
Q
=
2
\int_0^{\chi_\mathrm{p}^\mathrm{inj}}
p(\chi_\mathrm{p}) \, \mathrm{d}\chi_\mathrm{p} -1 \in[-1,1],
\label{eq:quantile}
\end{equation}
where $p(\chi_\mathrm{p})$ is the posterior distribution and $\chi_\mathrm{p}^\mathrm{inj}$ is the true value. The ideal case where the median of $p(\chi_\mathrm{p})$ coincides with the true value corresponds to $Q=0$. Obtaining $Q>0$ ($Q<0$) instead implies that the amount of precession in the   system is being overestimated (underestimated), %
and $100|Q|<X$ implies that the injected value is inside the $X\%$ symmetric confidence interval of the recovered posterior.
The values of $Q$ for our 100 injections are shown in Fig.~\ref{fig:quantiles}. We find a strong decreasing trend of $Q$ for increasing values of $\chi_\mathrm{p}$, which becomes particularly evident in the $\chi_{\rm p}>1$ region. %
Our analysis indicates that, in general, statistical errors cause an underestimate of $\chi_{\rm p}$ whenever $\chi_{\rm p}>1$. 
In other words, given a waveform model, sources with two-spin precession require larger SNR for accurate measurement~{(cf. \cite{2014PhRvL.112y1101V,2016PhRvD..93h4042P})}.

With a completeness of 86.5\% and a contamination of 1.2\%, our results are, overall,  are extremely encouraging.\footnote{As common in binary classification \cite{2019sdmm.book.....I}, we define completeness = true positives / (true positives + false negatives) and contamination = false positives / (true positives + false positives).} The broader conclusion  is that sources with $\chi_{\rm p}>1$ and sufficiently high SNR $\rho\gtrsim 20$ in O4 can, in principle, be correctly identified as affected by two precessing spins.

{\normalsize {\Large }}
\subsection{Impact of the prior}
\label{subsec:outliers}

Our parameter-space exploration highlights a generic tendency to underestimate precession effects whenever $\chi_{\rm p}>1$. %
The steep feature at $\chi_{\rm p}\approx1$ shown in Fig.~\ref{fig:prior} strongly suggests that  this statistical bias is driven by the employed prior.  
To verify this, we select two injections among the 100 we have just presented with
posterior quantile $Q\approx-1$, i.e. where the displacement between the injected and recovered values of $\chi_\mathrm{p}$ is maximized. More specifically, the two systems we consider have
$\chi_\mathrm{p}\in\{0.48,1.92\}$, $Q\in\{-0.98,-1\}$, and $\rho\in\{32.8,23.6\}$, respectively.

In Fig.~\ref{fig:outliers} we compare the posterior distributions obtained under the standard uniformative prior as in Sec.~\ref{sub:uniformchip} against
additional inference runs
where we instead take a volumetric prior on the spins. The latter choice enhances the prior weight assigned to configurations with large spins (cf. Fig.~\ref{fig:prior}). 
For the injection with $\chi_{\rm p}=0.48$ (left panel), we recover $\chi_\mathrm{p}=0.20\substack{+0.22 \\ -0.16}$ with the standard prior and  $\chi_\mathrm{p}=0.38\substack{+0.14 \\ -0.16}$ with the volumetric prior. This example shows that
simply
changing the prior to an alternative that is equally well motivated ---why should vectors like the spins be distributed uniformly in magnitude instead of volume?--- can significantly mitigate the inferred bias. In this case, the  posterior quantile increases from $Q=-0.98$ to $Q=-0.75$. %
The improvement is less evident, but still present, for the injection with $\chi_\mathrm{p}=1.92$, where a volumetric prior yields $\chi_\mathrm{p}=1.28\substack{+0.37 \\ -0.49}$ compared to $\chi_\mathrm{p}=1.10\substack{+0.46 \\ -0.49}$ for the standard prior.

Figure~\ref{fig:outliers} also shows additional runs where the same sources are considered at higher SNR, larger by a factor
$\approx3$,
using the standard priors.
 As already shown in Sec.~\ref{sub:differentSNR}, the posterior tends toward the true values for louder sources. %
Prior effects are still evident, with the true value remaining outside the 90\% credible interval. %
More specifically, for these high-SNR runs we find $\chi_\mathrm{p}=0.42\substack{+0.04 \\ -0.05}$ and $\chi_\mathrm{p}=1.71\substack{+0.14 \\ -0.19}$ for the $\chi_{\rm p}=0.48$ and $\chi_{\rm p}=1.92$   %
 case, respectively.

\begin{figure*}
    \includegraphics[width=0.9\textwidth]{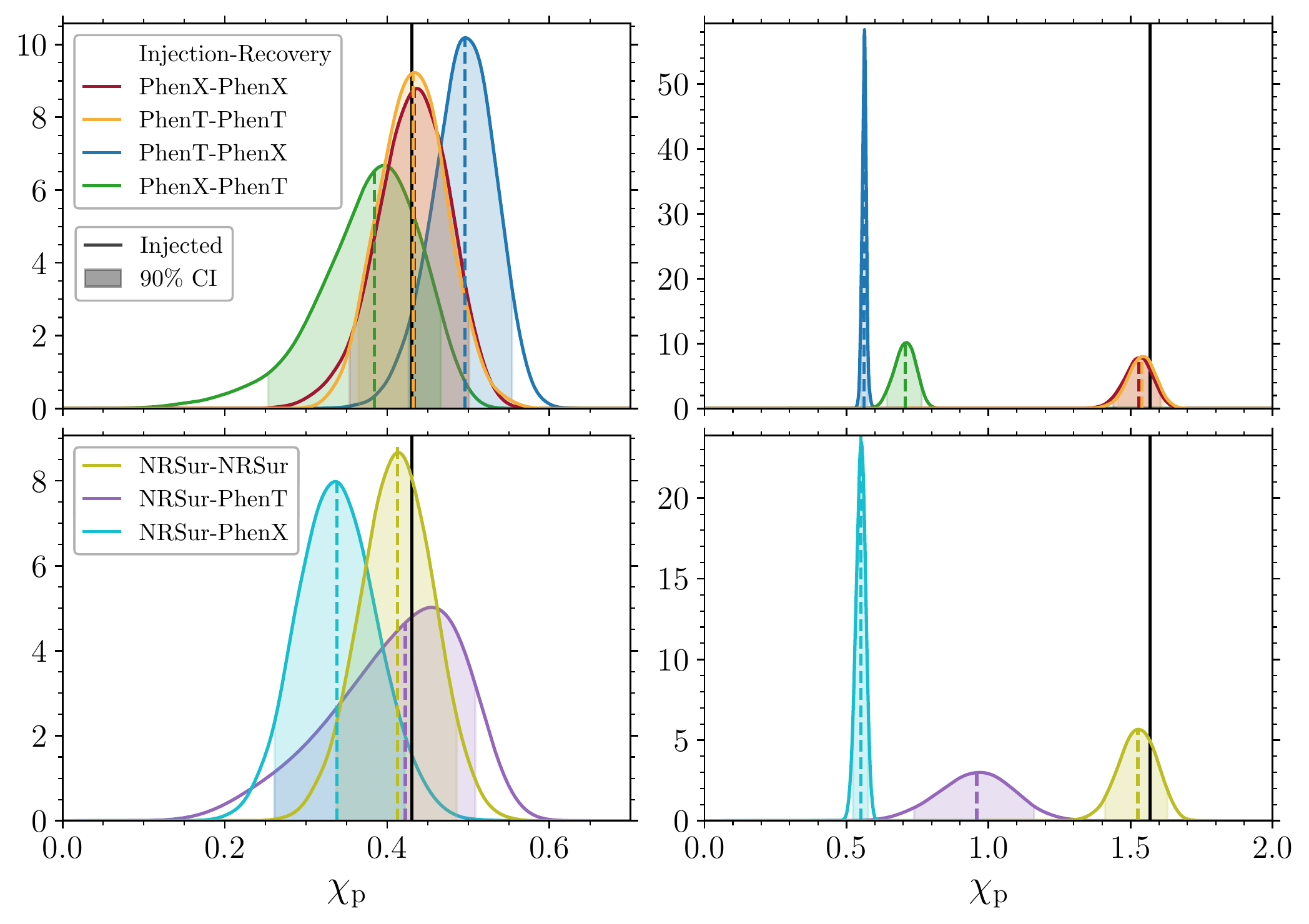}
    \caption{Posterior distributions of the averaged $\chi_\mathrm{p}$ parameter obtained with seven different combinations of waveform models.  The left (right) panels show results for a source injected with $\chi_{\rm p}=0.43$ and $M=131.1M_{\odot}$ ($\chi_\mathrm{p}=1.57$ and $M=130.8M_{\odot}$) %
    The top panels show results obtained with the two phenomenological models  {\sc IMRPhenomXPHM} (``PhenX'') for $\chi_\mathrm{p}=0.43$ ($\rho=107.3$) and for $\chi_\mathrm{p}=1.57$ ($\rho=90.3$) and  {\sc IMRPhenomTPHM} (``PhenT'') for $\chi_\mathrm{p}=0.43$ ($\rho=93.6$) and $\chi_\mathrm{p}=1.57$ ($\rho=81.7$). The bottom panels show results obtained in combination with the numerical-relativity surrogate model {\sc NRSur7dq4} (``NRSur'') for $\chi_\mathrm{p}=0.43$ ($\rho=100.2$) and for $\chi_\mathrm{p}=1.57$ ($\rho=75.6$). For each case, the label reported before (after) the hyphen in the legend refers to the waveform model used at the injection (recovery) stage. The injected values are indicated with black vertical lines. Medians and 90\% credible interval of the posterior distribution are indicated with dashed lines and shaded areas, respectively.
}
\label{fig:WF}
\end{figure*}

\subsection{Waveform systematics}
\label{sec:WF_systematics}

All the analyses illustrated so far were performed using the same waveform model for both injection and recovery and, therefore, do not capture systematic errors due to any mismodeling of the signal. Binaries with prominent spin effects are harder to model, implying that the $\chi_{\rm p}>1$ region we are interested in is also where discrepancies between different approximants are more likely to appear. %

Figure~\ref{fig:WF} and Table~\ref{tab:WF2} illustrate the posterior distribution of the averaged $\chi_\mathrm{p}$ parameter when different
models
are used in injection and recovery. We test various combinations of the {\sc IMRPhenomXPHM} \cite{2021PhRvD.103j4056P}, {\sc IMRPhenomTPHM}~\cite{2022PhRvD.105h4040E}, and {\sc NRSur7dq4}~\cite{2019PhRvR...1c3015V} waveform models.
We concentrate on two systems selected from the 100 injections presented in Sec.~\ref{sub:uniformchip}. %
In particular, we consider one source with $\chi_\mathrm{p}=1.57$ characterized by two prominently precessing spins as well as a control case with $\chi_\mathrm{p}=0.43$. The SNRs computed using {\sc IMRPhenomXPHM} are $\rho=90.3$ and $\rho=107.3$, respectively.  %
Both sources have sufficiently large detector-frame total masses $M\gtrsim 125 M_\odot$ such that the signal is short enough to be simulated with {\sc NRSur7dq4}. 
The cases where the signal is injected with {\sc NRSur7dq4} are arguably more realistic as this model is proven to be more accurate, i.e., closer to numerical-relativity simulations~\cite{2019PhRvR...1c3015V},
though the model does require extrapolation at the low-$q$ and high-$\chi_{\rm 1,2}$ edges of the parameter space we consider. %
For both the analyzed cases, the posteriors are relatively well centered on the true values whenever the injection and recovery are performed with the same waveform model.

The most evident  feature from Fig.~\ref{fig:WF} is that systematic biases increase dramatically for higher values of $\chi_\mathrm{p}$. %
This statement holds even though our analyzed low-$\chi_\mathrm{p}$ (high-$\chi_\mathrm{p}$) case has a higher (smaller) SNR and should thus be more (less) susceptible to waveform systematics. 

For the $\chi_\mathrm{p}=0.43$ source (left panels in Fig.~\ref{fig:WF}), the injected value lies inside the $90\%$ credible interval of the posterior for most the waveform combinations we tested. The only exception is the case where we inject with the  {\sc NRSur7dq4} and recover with {\sc IMRPhenomXPHM}. This run shows the largest quantile $Q=-0.93$ which tentatively suggests a lower accuracy of that model to spin precession, at least for this specific set of parameters. %
This conclusion is reinforced by our results obtained when the source is generated with {\sc IMRPhenomTPHM} but recovered with {\sc IMRPhenomXPHM}: the injected value is barely inside the 90\% credible interval with posterior quantile is $Q=-0.88$.

For the second case studied here with $\chi_\mathrm{p}=1.57$ (right panels in Fig.~\ref{fig:WF}), waveform systematics are severe. All waveform combinations where we inject and recover with different models return posterior distributions that are inconsistent with the true value at extremely high confidence (so high that we cannot meaningfully quantify it with the samples at our disposal). %
The worst cases appears to be those when we recover with {\sc IMRPhenomXPHM}, which produce a $\chi_\mathrm{p}$ posterior that is entirely below unity.

\begin{table}
\renewcommand{\arraystretch}{1.4}

  \begin{tabular}{cc||cc|cc|cc}
   
  &  \multirow{2}{*}{$\chi_\mathrm{p}^\mathrm{inj}=0.43$} &
      \multicolumn{2}{c|}{{PhenX}} &
      \multicolumn{2}{c|}{{PhenT}} &
      \multicolumn{2}{c}{{NRSur}} \\
  &    & {\textit{Q}} & {$\chi_\mathrm{p}$} & {\textit{Q}} & {$\chi_\mathrm{p}$} & {\textit{Q}} & {$\chi_\mathrm{p}$} \\
      \hline
  &  {PhenX} &0.06 & $0.43\substack{+0.07 \\ -0.08} $ &-0.6 & $0.38\substack{+0.08 \\ -0.13} $ &-- & --\\
  &  {PhenT} & 0.88 & $0.5\substack{+0.06 \\ -0.07}$ & 0.04 & $0.43\substack{+0.07 \\ -0.07}$ &-- & --\\
  &  {NRSur} & -0.93 & $0.34\substack{+0.08\\ -0.08} $  &-0.08 & $0.42\substack{+0.09 \\ -0.16} $ & -0.3 & $0.41\substack{+0.07\\ -0.08} $ \\
\\
  &      \multirow{2}{*}{$\chi_\mathrm{p}^\mathrm{inj}=1.57$} &
      \multicolumn{2}{c|}{{PhenX}} &
      \multicolumn{2}{c|}{{PhenT}} &
      \multicolumn{2}{c}{{NRSur}} \\
 &     & {\textit{Q}} & {$\chi_\mathrm{p}$} & {\textit{Q}} & {$\chi_\mathrm{p}$} & {\textit{Q}} & {$\chi_\mathrm{p}$} \\
 \hline
 &   {PhenX} &-0.6 & $1.53\substack{+0.08\\ -0.09} $ & -1 & $0.71\substack{+0.06\\ -0.07} $ &-- & --\\
 &   {PhenT} & -1 & $0.56\substack{+0.01 \\ -0.01} $ & -0.47 & $1.54\substack{+0.08 \\ -0.08} $ &-- & --\\     \rule[-0.4cm]{0mm}{0.cm}
 &   {NRSur} & -1 & $0.55\substack{+0.03 \\ -0.03}$ & -1 & $0.96\substack{+0.2 \\ -0.22}$& -0.48 & $1.53\substack{+0.1 \\ -0.12} $\\
  \end{tabular}
  
  \caption{Posterior quantiles $Q$, medians, and $90\%$ credible intervals of the averaged $\chi_{\rm p}$ estimator from analyses performed with three different waveform models: {\sc IMRPhenomXPHM} (``PhenX''),  {\sc IMRPhenomTPHM} (``PhenT''), and {\sc NRSur7dq4} (``NRSur''). The top (bottom) table shows results for injections with $\chi_\mathrm{p}=0.43$ ($\chi_\mathrm{p}=1.57$). In each table, the rows (columns) indicate the waveform used for signal injection (recovery). }
\label{tab:WF2}
\end{table}

While a more complete investigation on waveform systematics is beyond the scope of this work, the selected cases studied here tentatively indicate that current state-of-the-art approximants struggle at providing a consistent modeling of the signal in the $\chi_{\rm p}>1$ region, to a level that will be significant for the heavy, loud sources expected in O4. The discrepancies between the waveform models in the high-precession limit are potentially expected due to the differences in the prescriptions for the spin dynamics between the models. In {\sc IMRPhenomXPHM}, the precession angles are calculated by applying the stationary phase approximation to the multi-timescale scale analysis of the precession equations \cite{2015PhRvD..92f4016G,2017PhRvD..95j4004C}. The prescription is then artificially extended through the merger and ringdown beyond its regime of validity. In {\sc IMRPhenomTPHM}, the precession angles are calculated by direct integration of the equations of motion, coupled to a semianalytical approximation for the merger-ringdown that relies on an angular velocity determined by the quasinormal mode frequencies of the remnant BH \cite{2018arXiv180610734M}.

Waveform developers are actively working toward calibrating PN-based waveform models using numerical-relativity simulations with precessing spins (e.g.,~\cite{2021PhRvD.104l4027H}),
which will hopefully alleviate the systematic deviations reported here. It is also important to stress that, for this exercise, we had to select sources with high-enough mass such that the signal is fully covered by {\sc NRSur7dq4}, but these are also the systems where the precession signature is expected to be weaker. This is because precession cycles are contained in the low-frequency part of the signal that gradually falls out of band as the total mass increases.

On a more positive note, the conclusion that emerges here is that waveform systematic do not produce false positive: if a future observation will deliver $\chi_{\rm p}>1$ at high confidence, it appears safe to claim that the BH binary had two precessing spin.

\section{Conclusions}
\label{sec:conclusions}

If an incoming LIGO/Virgo source  is composed of merging BHs with two precessing spins, will we able to tell? In this paper we have provided a statistical assessment of this question using a large set of software injections. %

For dimensionality reduction and interpretation purposes, it is useful to have a single parameter that can capture the effect of precession in GW data. We employ a recent generalization \cite{2021PhRvD.103f4067G} of the effective precessing spin $\chi_\mathrm{p}$~\cite{2015PhRvD..91b4043S}. %
Unlike its predecessor, the augmented formulation does not assume that one of the two spin dominates the dynamics. %
In particular, the region of the parameter space
 $1<\chi_\mathrm{p}\leq 2$ is exclusive to binaries with two precessing spins. %
Because spin-spin couplings in GW data provide a weak contribution to the waveform, measuring {a} source with $\chi_{\rm p}>1$ also requires sensitive detectors. While such a detection has not occurred in current data~\cite{2021PhRvD.103f4067G,2022CQGra..39l5003H}, %
our software injections at O4 sensitivity demonstrate  that this goal is well within our reach.

\begin{figure}
\centering
    \includegraphics[width=\columnwidth]{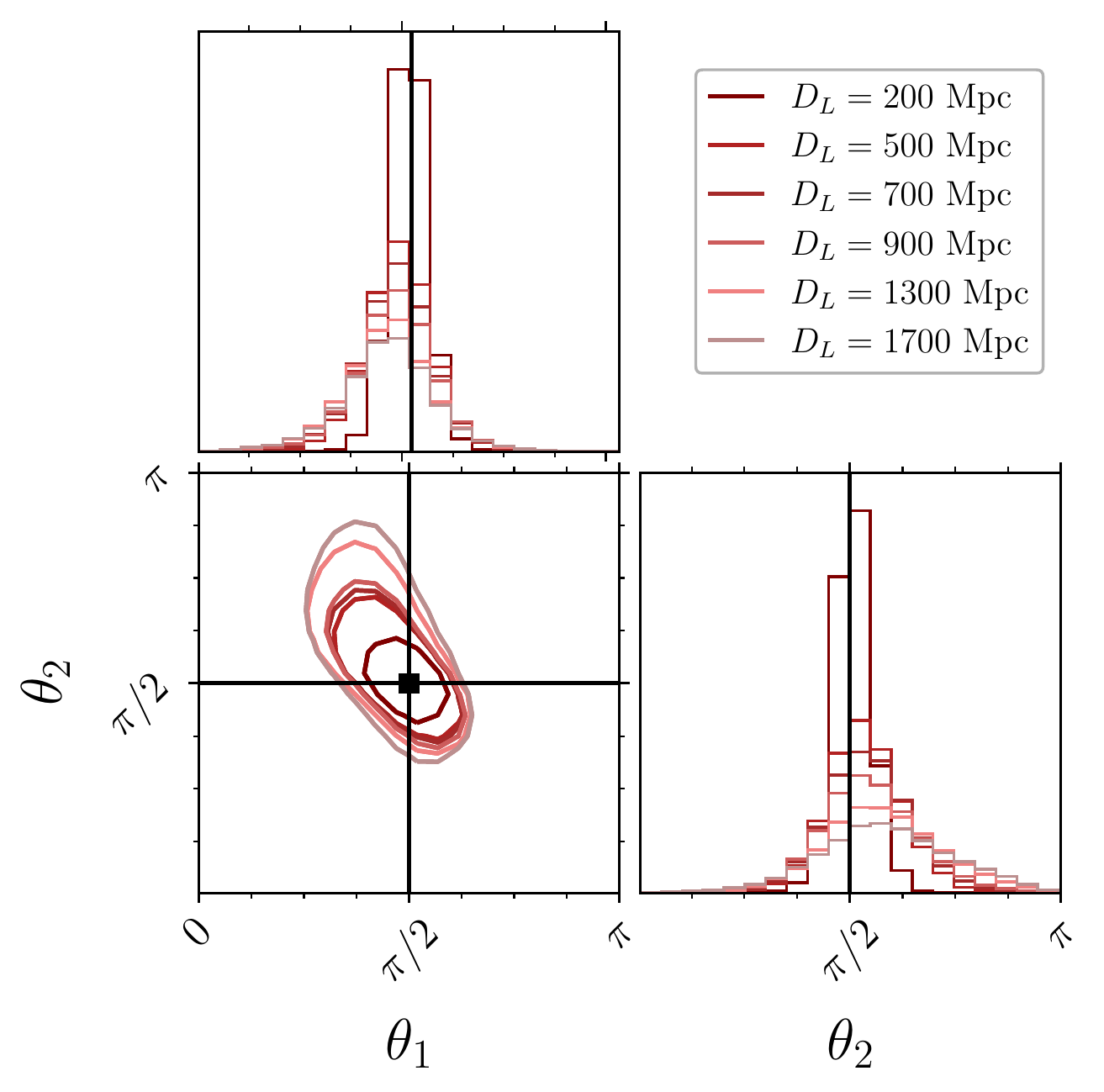}
    \caption{Joint posterior distribution of the tilt angles $\theta_{1}$ and $\theta_{2}$ for the single-system series described in Sec.~\ref{sub:differentSNR}. Darker (lighter) curves and histograms refer to sources at smaller (larger) distance. Contour level correspond to 90\% credible intervals.}
    \label{fig:corner}
\end{figure}

In this paper, we have concentrated solely on inference of precession with the $\chi_{\rm p}$ estimator. With the BH-binary parameter space spanning 15 dimensions, our large set of injections naturally contains much more information that could potentially be extracted (including, but not limited to, different spin precession estimators, correlation between the effective spins and other binary parameters,
and the vector spin components themselves).
In order to facilitate further exploitation, our posterior chains are made publicly available in their entirety at \href{https://github.com/ViolaDeRenzis/twoprecessingspins}{github.com/ViolaDeRenzis/twoprecessingspins}~\cite{datarelease}. The total computational budget to collect these data amounts to about half a million CPU hours.%

It is important to stress that the occurrence of two precessing spins is a necessary, but not sufficient, condition to obtain values  $\chi_{\rm p}>1$,
(i.e., being in this region implies the source has two precessing spins, but not vice versa).
Such large values of $\chi_{\rm p}$ require a considerable fine-tuning of the binary's intrinsic parameters (large spin magnitudes, mass ratios close to unity, spins coplanar with the orbit and aligned with each other). This makes our assessment very conservative and lets us identify sources with smoking-gun evidence of two-spin precession.

As the detectors' sensitivities increase and one moves beyond effective-spin parametrizations, inference on the higher-dimensional spin parameter space will hopefully allow us to relax such conservative assumptions. Looking ahead in this direction, constraining a source away from the edges of the $\theta_{1}$--$ \theta_{2}$ plane can also be taken as a telltale sign of two-spin precession (recall that $0\leq \theta_{i} \leq \pi)$.
Figure~\ref{fig:corner} shows the joint posterior distribution of the spin tilts for the series of injections described in Sec.~\ref{sub:differentSNR} where, indeed, nonprecessing configurations can be largely excluded. One point worth stressing is the short-timescale dependence of the tilt angles, which results in deeper issues when performing population studies \cite{2022PhRvD.105b4076M}. In contrast, the formulation of $\chi_{\rm p}$ explored here only varies on the longer radiation-reaction timescale ---which is the best one can hope for in the absence of additional constant of motions besides $\chi_{\rm eff}$ \cite{2008PhRvD..78d4021R}. %

Our large-scale injection study shows that BH binaries with two prominently precessing spins  at sufficiently high SNR can be generically identified as such. We also pointed out how the departure from Gaussianity of the $\chi_{\rm p}$ distribution can be a precious indicator of the suboptimality of the adopted indicator.
At the same time, prior effects and waveform systematics introduce some interpretation issues that need to be further explored.

Of all the software injections we performed with different source parameters and waveform models, we  did not detect a single confident false positive (i.e., a source with $\chi_{\rm p}<1$ which is erroneously recovered as having $\chi_{\rm p}>1$). Our study strongly indicates that, should a confident detection with $\chi_{\rm p}>1$ be made in O4, this would provide a conservative and safe claim of the first observation of a merging BH binary with two precessing spins.

\acknowledgements

\vspace{-0.1cm}
 We thank Daria Gangardt, Nathan Steinle, Floor Broekgaarden, and Roberto Cotesta for discussions.
V.D.R., D.G., and M.M. are supported by European Union's H2020 ERC Starting Grant No. 945155-GWmining, Cariplo Foundation Grant No. 2021-0555, and Leverhulme Trust Grant No. RPG-2019-350. V.D.R. acknowledges support from H2020 project HPC-EUROPA3 (INFRAIA-2016-1-730897). G.P. and P.S. acknowledge support from STFC Grant No. ST/V005677/1.
Computational work was performed at CINECA with allocations through INFN, Bicocca, and ISCRA project HP10BEQ9JB. %
 This manuscript has LIGO document number P2200196.%

\bibliography{chipinjections}

\end{document}